\documentclass{JHEP3} 

\usepackage{epsfig,multicol}

\newcommand{\ds}{{\sffamily DarkSUSY}}
\def\msun{M_{\odot}{\ }}
\newcommand{\code}[1]{{\tt #1}}

\newcommand{\PLB}[3]{ Phys.~Lett. \textbf{B#1} (#2) #3}

\newcommand{\HPA}[3]{ Helv.~Phys.~Acta \textbf{#1} (#2) #3}

\newcommand\fverb{\setbox\pippobox=\hbox\bgroup\verb}
\newcommand\fverbdo{\egroup\medskip\noindent%
			\fbox{\unhbox\pippobox}\ }
\newcommand\fverbit{\egroup\item[\fbox{\unhbox\pippobox}]}
\newbox\pippobox
\newcommand{\beq}{\begin{equation}}
\newcommand{\eeq}{\end{equation}}

\def\neu1{\widetilde{\chi}^0_1}
\def\vf{\varphi}
\def\simlt{\stackrel{<}{{}_\sim}}
\def\simgt{\stackrel{>}{{}_\sim}}

\title{Electroweak Baryogenesis, Large Yukawas and Dark Matter}

\author{Alessio Provenza \\
	SISSA/ISAS, via Beirut 2-4, 34013 Trieste, Italy\\
	E-mail: \email{provenza@he.sissa.it}}

\author{Mariano Quiros \\
	Instituci\'o Catalana de Recerca i Estudis Avan\c{c}ats (ICREA) \\
        Theory Physics Group, IFAE/UAB,
        E-08193 Bellaterra, Barcelona, Spain \\
	E-mail: \email{quiros@ifae.es}}
	
\author{Piero Ullio \\
	SISSA/ISAS, via Beirut 2-4, 34013 Trieste, Italy\\
	E-mail: \email{ullio@sissa.it}}

\preprint{SISSA-58/2005/EP\\UAB-FT-585}

\abstract{It has recently been shown that the electroweak baryogenesis
mechanism is feasible in Standard Model extensions containing extra
fermions with large Yukawa couplings. We show here that the lightest
of these fermionic fields can naturally be a good candidate for cold
dark matter. We find regions in the parameter space where the thermal
relic abundance of this particle is compatible with the dark matter
density of the Universe as determined by the WMAP experiment. We study
direct and indirect dark matter detection for this model and compare
with current experimental limits and prospects for upcoming
experiments. We find, contrary to the standard lore, that indirect
detection searches are more promising than direct ones, and they
already exclude part of the parameter space.}
 

\begin{document} 

\section{Introduction}

The Standard Model (SM) of elementary particles is extremely accurate
in describing the fundamental interactions up to the energy scale
probed so far at accelerators. Its application to cosmology has led as
well to significant successes such as the prediction of the light
elements abundance. However the SM fails to provide a pattern to embed
all features emerging from recent data on precision cosmology: in
particular, neither it does provide a mechanism to explain the origin
of the matter-antimatter asymmetry in the Universe, nor it does
accommodate a candidate for non-baryonic cold dark matter.

Baryogenesis and the dark matter problem stand as two of the most
intriguing topics of research in today's Science, and they have been
examined at length from very different perspectives. It is interesting
to notice that, among other viable approaches, both issues have been
addressed invoking new physics at the weak scale, at an energy which
stands around the corner with respect to upcoming tests of fundamental
interactions at present and future accelerators. A successful
electroweak baryogenesis~\cite{ewb} can arise in SM extensions in
which new particles make the electroweak phase transition strongly
first order and add new sources of CP-violation. On the other hand
weakly interacting massive particles (WIMPs) are among the leading
candidates for dark matter~\cite{wimps}.  All these ingredients can be
provided in a single framework, as it is the case of supersymmetric
extensions of the SM. Even within the minimal supersymmetric SM
(MSSM), if the (mostly) right-handed stop is lighter than the top
quark and the Higgs is sufficiently light, electroweak baryogenesis
may be realized~\cite{CQW} and, at the same time, if the lightest
supersymmetric particle is a neutralino, it can play the role of WIMP
dark matter candidate.

In a recent paper Carena {\it et al.}~\cite{Carenaetal} have shown
that in order to strengthen the electroweak phase transition it is not
strictly required to consider models with light extra bosonic degrees
of freedom, as it was the case in all electroweak baryogenesis models
considered in the past, but that models with extra fermions can be
equally successful provided that large Yukawa couplings are
introduced. A simple implementation of this idea involves introducing
doublet and triplet fermions, such as {\em e.g.}~Higgsinos and gauginos,
which can carry as dowry new charge- and color-neutral particles, the
lightest of which can be the dark matter candidate. Carena {\it et
al.}~discuss in details one such simple setup, reminding in some
aspects split supersymmetry~\cite{split}, and explicitly show that it
can indeed provide electroweak baryogenesis.

In this article we discuss the dark matter features of the above
model, as well as those of a slightly extended framework model.
Although here, as in the MSSM, the dark matter candidate is a
neutralino, the physical state from the superposition of two gaugino
and two Higgsino fields, we point out that there are significant
differences compared to the MSSM case, both in the mechanism setting
its thermal relic density and in its phenomenology as dark matter
candidate. In particular we show that currently the tightest
constraint on the model comes from limits on the neutrino induced flux
from pair annihilation of dark matter neutralinos gravitationally
trapped in the center of the Sun. We also show that the best option
for the upcoming future is to measure an excess in antimatter cosmic
ray fluxes, while direct detection in underground laboratories looks
less promising.

The plan of the paper is as follows. In Section~\ref{sec:fram} we
review the framework introduced in Ref.~\cite{Carenaetal} and discuss
relevant features from the perspective of the dark matter problem. In
Section~\ref{sec:guide} we discuss how dark matter candidates arise in
a model with a reduced number of parameters, for which the
baryogenesis mechanism was discussed in detail in
Ref.~\cite{Carenaetal}. In Section~\ref{sec:phenoguide} we present
limits on the model from current searches and future perspectives,
going through all the techniques of WIMP dark matter detection. In
Section~\ref{sec:towards} we discuss the dark matter thermal relic
density and current and future searches in a more generic model,
within the framework discussed in detail in Ref.~\cite{Carenaetal},
where future direct detection is more promising. Finally our
conclusions are drawn in Section~\ref{sec:conclusion}.

\section{Particle physics framework}
\label{sec:fram}

We consider a minimal extension of the SM in which the role of the
extra fermions with large Yukawa couplings is played by fields with
the quantum numbers of Higgsinos and gauginos in supersymmetric
theories; such fields are assumed to be the only light extra particles
present in our theory and relevant for its weak-scale
phenomenology. Our extra fields have gauge interactions as ordinary
Higgsinos and gauginos in the MSSM, while we define their couplings to
the SM Higgs doublet $H$ through the Lagrangian:
\begin{eqnarray}
\mathcal L&=&H^\dagger\left(h_2\,\sigma_a\tilde W^a+h^\prime_2\, \tilde
B\right)\tilde H_2 +H^T\epsilon\left(-h_1\,\sigma_a\tilde
W^a+h^\prime_1\, \tilde B\right)\tilde H_1\nonumber\\
&+&\frac{M_2}{2}\,\tilde W^a\tilde W^a
+\frac{M_1}{2}\, \tilde B\tilde B+\mu\, \tilde H^T_2\epsilon \tilde H_1+
h.c.\;,
\end{eqnarray}
with $\epsilon = i\,\sigma_2$. The setup we have introduced has a
particle content analogous to the split supersymmetry scenario, a MSSM
in which all scalars, except for the SM-like Higgs, are driven at a
very heavy mass scale. Hence one could regard it as a particular
realization of split supersymmetry in which the standard relation
between gauge and Yukawa couplings has been spoiled~\footnote{See
Ref.~\cite{Carenaetal} for a discussion on sample realizations of this
model as a low energy effective limit in supersymmetric theories.}.
Another difference being that in split supersymmetry the gluino may be
light, with relevant phenomenological implications but unrelated to
the dark matter or baryogenesis problems at focus here. For reference,
ordinary MSSM couplings are recovered if the generic Yukawa couplings
$h_{1,2}$ and $h^\prime_{1,2}$ are chosen as:
\begin{eqnarray}
 h_1=g\cos\beta/\sqrt{2} & \hspace{1.5cm} & h_2=g\sin\beta/\sqrt{2}
 \nonumber \\ h_1^\prime=g^\prime\cos\beta/\sqrt{2} &
 &h_2^\prime=g^\prime\sin\beta/\sqrt{2}
\end{eqnarray} 
where $g$ and $g^\prime$ are the $SU(2)$ and $U(1)$ gauge couplings.
Modifications to these relations can appear in other non-minimal
contexts. Keeping this in mind, we will take the Yukawa couplings as
free parameters and we will discuss the phenomenology of the model
regardless of its eventual supersymmetric completion at high energy.

Four physical neutral states $\tilde{\chi}^0_i$ and two physical
charged states $\tilde{\chi}^{\pm}_i$ are obtained by diagonalizing
the corresponding mass matrices.  According to our definitions, in the
basis $\left(\tilde B^0, \tilde W^0, \tilde H^0_1, \tilde H^0_2
\right)$, the neutralino mass matrix takes the form:
\begin{equation}
\left(
\begin{array}{cccc}
M_{1} & 0 & - \sqrt{2}\,h_{1}^{\prime }\,m_{W} /g &
\sqrt{2}\,h_{2}^{\prime }\,m_{W} /g \\ 0 & M_{2} &
\sqrt{2}\,h_{1}\,m_{W} /g & - \sqrt{2}\,h_{2}\,m_{W} /g \\ -
\sqrt{2}\,h_{1}^{\prime }\,m_{W} /g & \sqrt{2}\,h_{1}\,m_{W} /g & 0 &
-\mu \\ \sqrt{2}\,h_{2}^{\prime }\,m_{W} /g & - \sqrt{2}\,h_{2}\,m_{W}
/g & -\mu & 0
\end{array}
\right)\;\label{neumatrix}
\end{equation}
while the chargino mass matrix is:
\begin{equation}
\left(
\begin{array}{cc}
M_{2}                     & 2\,h_{2}\,m_{W} /g  \\
2\,h_{1}\,m_{W} /g &   \mu  
\end{array}
\right)\;.
\end{equation}
Since baryogenesis stands as the main motivation of our framework, we
need to introduce a non-vanishing CP-violating phase, triggering
baryon number generation: as a minimal assumption, it is sufficient to
take the Higgsino mass parameter to be complex, $\mu = |\mu|
\,e^{i\vf}$, while choosing the gaugino mass parameters $M_1$ and
$M_2$, and the Yukawa couplings to be real.

\subsection{Lightest neutralino mass and composition}

We focus on the case in which the lightest neutralino (LN)
$\tilde{\chi}^0_1$ is the lightest extra-fermion (LEF) and hence a
stable species~\footnote{Contrary to the MSSM, here the lightest
chargino can be lighter than the lightest neutralino.}: the LN being
electric- and color-charge neutral, massive and stable it is an ideal
candidate for cold dark matter.  The phenomenology of the LN as a dark
matter candidate crucially depends on its mass and mixing; in
particular, the relative weight between its gaugino and Higgsino
components is decisive both in setting the LN thermal relic abundance
and in determining the detection prospects of such a model.

The gaugino or Higgsino nature of the LN is related as usual to the
hierarchy among the parameters $M_1$, $M_2$ and $\mu$. Since in our
model there are significant differences compared to the most commonly
considered cases in the MSSM context, we preliminarily sketch here
some trends in a few sample cases in which it is possible to
diagonalize analytically the neutralino mass matrix.

As a first example we focus on the setup in which $M_1$ is very heavy,
{\em i.e.}~$|M_1| \gg |\mu|,|M_2|, \sqrt{2}\,h_{1,2}\,m_{W} /g$, and
hence the $\tilde B^0$ component decouples. For simplicity we also
assume that $\mu$ is real and $h_1=h_2 \equiv h$~\footnote{Obviously,
the values of $h_{1}^{\prime }$ and $h_{2}^{\prime }$ do not play any
role here.}. In this case the three light eigenvalues of the
neutralino mass matrix are:
\begin{equation}
\lambda_{\pm}={1\over2}\left(M_2+\mu\pm\sqrt{(M_2-\mu)^2+16 \,
h^2m_{W}^2/g^2}\right)\;, \;\;\;\;\;\; \lambda_3=\mu\;.
\end{equation}
In the limit $|\mu| \gg |M_2|$, the lightest eigenvalue is $\lambda_-
\simeq M_2-4 \, h^2\,m_{W}^2/g^2\mu$ and the associated eigenvector is
mostly Wino-like, with a Higgsino component induced by the Yukawa term
which gets smaller and smaller as the $M_2$ scale gets much larger
than $h\,m_{W}/g$ (recovering the limit one would have in the
MSSM). In the opposite regime $|\mu| \ll |M_2|$ there are two light
states: $\lambda_- \simeq \mu-4 \, h^2\,m_{W}^2/g^2M_2$ and $\lambda_3
= \mu$. If $\mu$ and $M_2$ have the same sign the first state is the
lightest one and the associated LN is mostly Higgsino-like, with a
Wino component again introduced by the Yukawa terms. If $\mu$ and
$M_2$ have opposite signs the state with mass $\lambda_3 $ becomes the
lightest one and now the LN is an almost pure Higgsino state.  These
three regimes and the corresponding LN compositions are schematically
summarized in the left panel of Fig.~\ref{plane}, in the plane
$(\mu;\,M_2)$ and for the sample value $h=1$.

\FIGURE[t]{
\centerline{
\epsfig{file=plot/binodec.eps,width=7.5cm}  \quad
\epsfig{file=plot/allequal.eps,width=7.5cm}} 
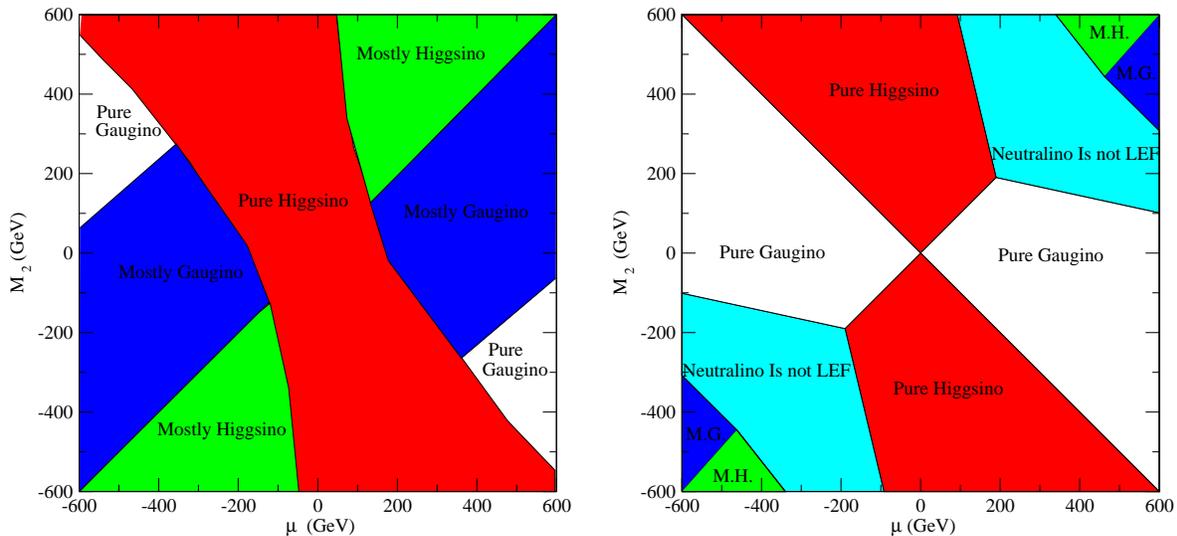
\caption{The gaugino and Higgsino contents of the lightest neutralino
in the plane ($\mu;\,M_2$).  We label pure Higgsino (gaugino) the
state with a Higgsino (gaugino) component greater than 90\%.  A mostly
Higgsino (gaugino) state has a Higgsino (gaugino) component between
50\% and 90\%. We assume Yukawa couplings $h=1$ as sample reference
value.  In the left panel $M_1$ has been fixed at a very heavy scale;
in the right panel we assumed $M_1 = M_2$.}
\label{plane}
}

Another simple case showing the interplay between mass parameters and
mixings can be built by choosing {\em e.g.}~$M_1=M_2=M$ and
$h_{1,2}=h_{1,2}^\prime=h$ (still $\mu$ is taken to be real).  In this
case the eigenvalues of the neutralino mass matrix are:
\begin{eqnarray}
\lambda_{\pm}& =
&{1\over2}\left(M+\mu\pm\sqrt{(M-\mu)^2+32h^2m_{W}^2/g^2}\right)\;,\nonumber
\\ \lambda_3&=&M\;, \;\;\;\;\;\; \lambda_4=\mu\;.
\end{eqnarray}
Again the state with mass $\lambda_4$ is almost pure Higgsino with
mass $\mu$, while that with mass $\lambda_3$ is an almost pure gaugino
state with mass $M$. Due to the Yukawa coupling this gaugino state is
a mixing between the Bino and the Wino in the same
percentage~\footnote{This Bino-Wino mixing is a peculiar feature of
our setup which never appears in the MSSM.}.  As in the previous
example the remaining states with masses $\lambda_\pm$ are
gaugino-Higgsino mixed states with relative weight depending on the
values of the parameters.  The hierarchy between eigenvalues is
analogous to the previous case, except that now the state with mass
$\lambda_-$ becomes the lightest one only in the case where both $\mu$
and $M$ are fairly large.  Regions of different LN compositions are
shown in Fig.~\ref{plane}, right-panel. Note there is a region in the
parameter space in which the LN is not the lightest extra fermion,
since a chargino becomes lighter.

Another case we will consider later on regards the possibility to fix
the ratio between the gaugino mass parameters, {\em i.e.}~$M_1= \alpha
M_2$.  The general trend here is very similar to the previous one,
taking into account that by varying the parameter $\alpha$ we change
the Bino content of the gaugino-like LN.  At the same time in the
limit $\alpha\gg 1$ one recovers the complete Bino decoupling as in
the first case discussed here, except for the corner at very small
$M_2$, where $M_1\gg h\,m_W/g$ does not hold any more and hence the
Bino does not decouple.

\section{Dark matter candidates in a guideline model}
\label{sec:guide}

Analogously to the MSSM with $R$-parity conservation, in our scenario
the lightest extra fermion $\neu1$ is stable, massive and weakly
interacting, and hence a natural WIMP candidate for cold dark matter
(CDM).  We compute the LN thermal relic density by interfacing the
particle physics framework we have introduced in the \ds\
package~\cite{Gondolo:2004sc}. Such package allows for high accuracy
solutions of the Boltzmann equations describing thermal freeze out.
In particular, in computing thermally-averaged LN pair annihilation
cross-sections~\footnote{The relic density is roughly speaking
proportional to its inverse.}, all kinematically allowed final states
are systematically included, as well as eventual co-annihilation
effects~\footnote{In case there are extra particles nearly degenerate
in mass with the LN, such initial states, properly weighted, should be
included too.}. The density evolution equation is then solved
numerically.  The estimated precision on the value of the relic
density we derive is, for a given set of input parameters setting
masses, widths and couplings, of the order of 1\% or better.  The LN
relic density value is to be compared with the latest determination of
the CDM component of the Universe by the WMAP
experiment~\cite{Spergel:2003cb}: $\Omega_{CDM} h^2 = 0.113 \pm
0.009$.

Our first working model within the framework will be that with a
reduced number of parameters discussed at length in
Ref.~\cite{Carenaetal} in the electroweak baryogenesis context.  We
first take the limit of Bino decoupling setting $|M_1| \gg
|\mu|,|M_2|$ and $h_{1,2}^{\prime} =0$, and then fix $\mu = - M_2
\,e^{i\vf}$, a condition which maximizes the number of degrees of
freedom contributing to strengthening the electroweak phase
transition. For this particular model it has also been explicitly
shown that one can build an ultraviolet completion canceling out
instabilities in the zero temperature Higgs potential induced by the
light extra fermions we have introduced~\footnote{In
Ref.~\cite{Carenaetal} heavier bosons coupled to the Higgs were
introduced in order to stabilize the effective potential. If these
heavy bosons are SM singlets they do not perturb the electroweak
observables nor they interfere the annihilation and detection rates of
the dark matter candidate.}. Our guideline model is then defined by
only five free parameters: $|\mu|$, $\vf$, $h_{+}=\frac{1}{ 2}\left(
h_{1} + h_{2}\right)$, $h_{-}=\frac{1}{ 2}\left( h_{1} - h_{2}\right)$
and the SM Higgs mass $m_H$. Going back to the list of limiting cases
we discussed in the previous Section, we see that we are referring to
a model in which the LN is an almost pure Higgsino with mass $M_{LN}
\simeq |\mu|$.

\FIGURE[t]{
\centerline{
\epsfig{file=plot/livello.eps,width=8.0cm}} 
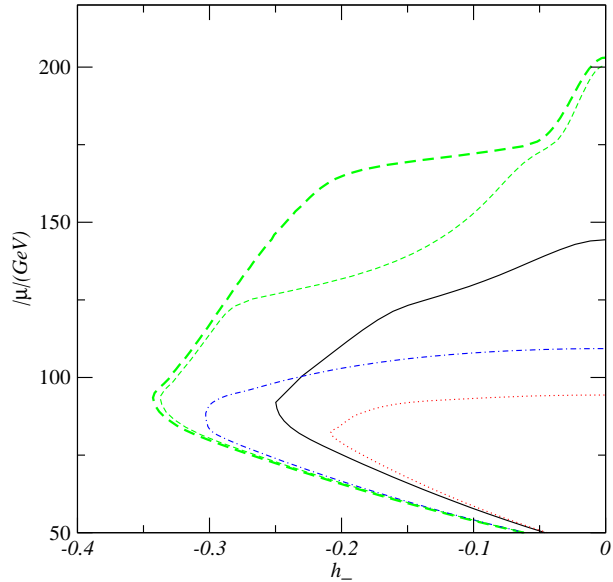
\caption{Sample isolevel curves of the lightest neutralino relic
abundance at the currently preferred value $\Omega_{LN} h^2 = 0.113$.
From top to bottom the extra free parameters are set equal to:
$h_+=2$, $\vf=0$ and $m_H=300$~GeV (thick dashed curve), $h_+=2$,
$\vf=0$ and $m_H=150$~GeV (thin dashed curve), $h_+=1.5$, $\vf=0$ and
$m_H=150$~GeV (solid curve), $h_+=2$, $\vf= \pi/2$ and $m_H=150$~GeV
(dash-dotted curve), and $h_+=1.5$, $\vf=\pi/2$ and $m_H=150$~GeV
(dotted curve). }
\label{fig:relic}}

In Fig.~\ref{fig:relic} we show isolevel curves at $\Omega_{LN} h^2 =
0.113$ in the plane $(|\mu|;\,h_-)$, for a few sample values of $\vf$,
$h_+$ and $m_H$. The relic density is sensitive to the parameter
$|\mu|$ since it drives the mass scales of the extra particles we have
introduced. We are restricting ourselves to the case $h_- < 0$, since
the model is symmetric under the exchange $h_- \rightarrow - h_-$. The
value of the relic density rapidly changes with $h_-$ because, as
pointed out in Ref.~\cite{Carenaetal}, the coupling $Z^0\neu1\neu1$ is
proportional to:
\begin{equation}
g_{Z^0 \neu1\neu1}\propto \frac{h_2^2-h_1^2}{h_2^2+h_1^2} = -
\frac{1}{2} \frac{h_+\, h_-}{h_-^2+h_+^2}\,.
\label{eq:zcoup}
\end{equation}
At values of the $\neu1$ mass below the threshold for pair
annihilation into gauge boson final states, the only open channel is
the helicity suppressed fermion-antifermion state, which gets its
largest contribution from the diagram with a $Z$ boson in the
$s$-channel. The annihilation rate gets maximal on resonance, at
$m_{\neu1}=m_Z/2$, driving the relic density to very small values
unless one considers a tiny $h_-$. Moving away from the resonance,
both toward heavier and lighter masses, the isolevel curves spread
out to larger and larger values of $h_-$. We only display in
Fig.~\ref{fig:relic} the upper branch since, for LN masses smaller
than $m_Z/2$, the induced contribution to the $Z$ invisible width is
larger than the experimental upper bound and such models are excluded
by LEP results.  For LN masses approaching the $W$ and $Z$ masses, LN
pair annihilations into $W^+\,W^-$ and $Z^0 \,Z^0$ in the early
Universe become relevant and tend eventually to dominate.  These
processes proceed mainly through $t$- and $u$-channel exchanges of,
respectively, the lightest chargino (LC) and the next-to-lightest
neutralino (NLN).

At first sight this picture may just seem the analogue to the
well-studied case of Higgsinos in the MSSM; there are however
substantial differences. In the MSSM, in the case of a pure Higgsino
LN, the lightest chargino and the next-to-lightest neutralino are
quasi-degenerate in mass with the LN: the contribution of the
$W^+\,W^-$ and $Z^0 \,Z^0$ final states to the cross-section becomes
too large as soon as the LN mass gets above the corresponding
threshold, driving the relic abundance to very small
values~\footnote{The mass degeneracy implies as well that large
co-annihilation effects appear, with a further reduction in the
lightest supersymmetric particle (LSP) relic abundance.}. To
compensate for this, one should increase the Higgsino LSP mass up to
the TeV range.

\FIGURE[t]{
\centerline{
\epsfig{file=plot/split.eps,width=7.1cm}\quad 
\epsfig{file=plot/crossw.eps,width=7.3cm}}
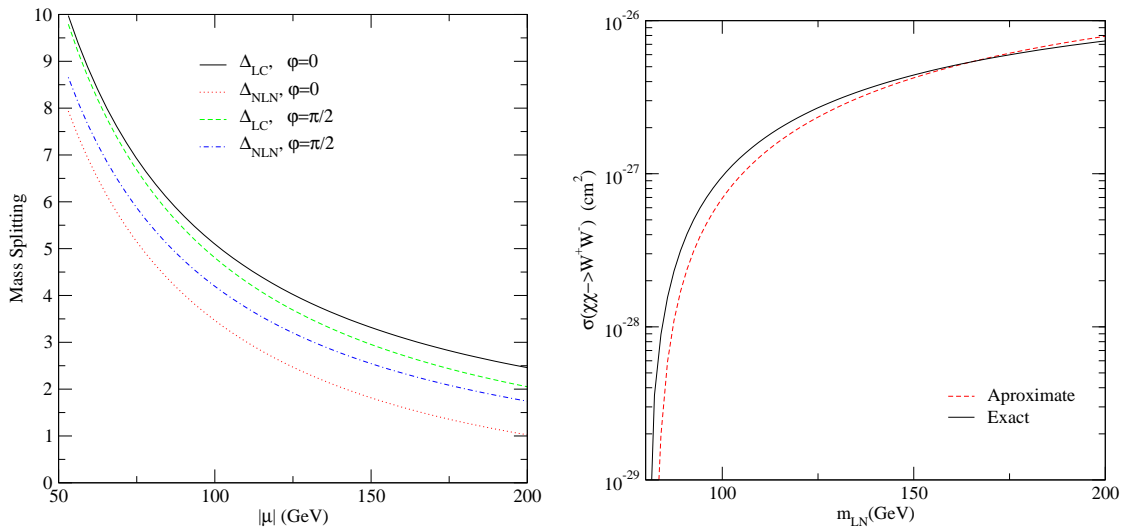
\caption{Left panel: The mass splitting between the lightest
neutralino (LN) and the lightest chargino (LC) or the next-to-lightest
neutralino (NLN), defined as
$\Delta_{i}=\left(m_{i}-m_{LN}\right)/{m_{LN}}$, as function of
$\mu$. We assumed $h_+=2$, with $\vf$ between extrema, and chosen
$h_-=-0.25$ (there is a mild shift in the mass splittings varying
$h_-$ in the interesting range for the relic abundance).  Right panel:
Cross-section for LN pair annihilation into a $W$-boson pair, in the
limit of particles in the initial state at rest, as function of the
lightest neutralino mass. The solid line corresponds to the
computation made with \ds , the dashed line to the result assuming
pure Higgsino couplings and a mass of the lightest chargino equal to
300 GeV (average value of $m_{LC}$ in the sample parameter choice).}
\label{fig:shift}}

On the other hand, in our framework, while $m_{\neu1}\simeq|\mu|$, the
mass scale of the other two fermionic states is instead set by the
largest of the Yukawa terms $h_1$ and $h_2$, or equivalently by $h_+$,
and the mass splittings can be very large. This is shown in the left
panel of Fig.~\ref{fig:shift}; the parameters $\vf$ and $|h_-|$
contribute as well in setting the mass splittings, but at a milder
level. It follows that, along the foliation of the parameter space we
are considering, at a fixed value of $h_+$ and varying $|\mu|$, the
rate of annihilation into gauge bosons increases with the LN mass
({\em i.e.}~with $|\mu|$), rather than decreasing with it as one would
intuitively expects and happens {\em e.g.}~in the MSSM. This trend is
shown for the $W^+\,W^-$ final state in the right panel of
Fig.~\ref{fig:shift}. The mass dependence we find for $\sigma v$
closely reproduces the scaling $m_{\neu1}^2/m_{LC}^4$ we have
displayed (dashed line) implementing the formula for the cross-section
for a pure Higgsino coupling and fixing $m_{LC}$ to the mean value in
our sample parameter choice.

Coming back to Fig.~\ref{fig:relic}, once above the threshold for
gauge boson production the isolevel curves bend back to the (small)
values of $|h_-|$ at which, for given $|\mu|$, $h_+$ and $\vf$ (and
hence the corresponding LN--LC and LN--NLN mass splittings), diagrams
with a chargino or neutralino exchange alone are large enough to drive
the relic abundance down to the WMAP range. At larger values of
$|h_-|$ the large coupling between LN and the $Z$ boson makes the $W$
boson final state to get an additional large contribution from the
diagram with the $Z$ boson in the $s$-channel. In each sample case the
region delimited by the isolevel curve and the vertical axis at $h_- =
0$ corresponds to relic densities larger than the central value from
the WMAP determination ({\em i.e.}~most of it is cosmologically
excluded), while at larger $h_-$ ({\em i.e.}~outside the region
delimited by the isolevel curves) relic densities are lower than the
value required for the LN to be the main dark matter
component~\footnote{Unless non-standard production mechanisms are
invoked, or non-standard cosmological setups implemented, two
possibilities that will not be further considered
here.}. Modifications to the general trend we sketched come from
eventual additional contributions when other final states become
kinematically allowed. In the examples with $h_+=2,\,\vf=0$, mass
splittings between the extra fermions are the largest ones and the
isolevel curves stretch for LN masses above the top mass: the process
$\neu1\neu1\rightarrow t\overline{t}$ comes into play and, since this
channel is not helicity suppressed, it gets large contributions
through the s-channel Z exchange down to very small values of $|h_-|$,
$|h_-|\simeq 5\times 10^{-2}$.  If the $Z^0 H$ final state becomes
kinematically allowed, it can give as well a substantial contribution
to the annihilation cross-section. This effect is shown again for the
two curves with $h_+=2,\,\vf=0$: one with $m_H=150$~GeV, {\em
i.e.}~close to the presently preferred value from electroweak
precision measurements~\cite{PDG}, the other with $m_H$ twice as
large. For $m_H=150$~GeV the $Z^0 H$ threshold opens up at $|\mu| \sim
120$~GeV: above it the two isolevel curves depart from each other,
while below it the curves essentially coincide since $H$ enters only
through contributions to s-channel diagrams, always much smaller than
the corresponding $Z^0$ s-channel diagram. In the same way, the shape
of the isolevel curves for $h_+=1.5,\,\vf=\pi/2$ and
$h_+=2,\,\vf=\pi/2$ remains essentially unchanged in case we rise the
value $m_H=150$~GeV to much larger values.

The sample cases we have considered, with fairly large
$h_+$~\footnote{But much smaller than the generic upper limit from the
requirement of perturbativity of the theory at low scale, about
$\sqrt{4 \pi}$.} and moderately light $m_H$ are among those found to
be favored by electroweak baryogenesis in Ref.~\cite{Carenaetal}, and
hence they are good cases to check the phenomenology of the dark
matter model in the framework where the baryogenesis problem is
simultaneously addressed. At the same time, the size of the
CP-violating phase has been varied freely between extremes, and will
enter in our discussion through the shift in mass splittings only. For
transparency in our discussion, and to present results which have a
validity on their own, we are not going to zoom in only sub-slices of
the parameter space which are fully successful in electroweak
baryogenesis. In the same vein, we implement as sharp cut to the
parameter space only the bound on extra contributions to the $Z$
invisible width. As it was discussed in Ref.~\cite{Carenaetal},
significant constraints on the model can be extracted also from
experimental bounds on the electroweak $T$-parameter: the slice of
models included in our plot at the largest value of $|h_-|$ are
excluded for small values of $m_H$, while the constraint gets weaker
going to a heavier Higgs. On the other hand, as it was shown in the
relic density calculation and discussed further below, the
phenomenology of the dark matter candidate is almost insensitive to
$m_H$, except for the threshold effect already described. We will keep
all models and show that actual current limits on dark matter searches
introduce even tighter constraints, independently on the value of
$m_H$.

\section{Detection rates in the guideline model}
\label{sec:phenoguide}

The issue of WIMP dark matter detection has been studied at length
(for reviews see {\em e.g.}~\cite{jkg,Bergstrom:1998xh}).  We will
systematically go through all WIMP detection techniques to illustrate
those that already exclude models within our framework and what are
the detection prospects for the future. As for the relic density, all
rates are computed with the \ds\ package~\cite{Gondolo:2004sc}.  The
set of underlying assumptions is briefly reviewed here, while present
limits and the future outreach is discussed within the simplified
framework of our guideline model.

\subsection{Direct detection}

In the last decade considerable resources have been invested in the
attempts to directly detect WIMPs, {\em i.e.}~to measure the energy
deposited in elastic scatterings off of nuclei by dark matter WIMPs
passing through the target material of a detector~\cite{dirdet}. We
present predictions in terms of scattering cross-sections on a single
nucleon, separating as usual the term accounting for coherent
spin-independent (SI) interactions from the one due to axial-vector
spin-dependent (SD) coupling. In our framework the process of
scattering of a LEF on a quark is particularly constrained since only
a $t$-channel exchange of the SM Higgs boson mediates the SI part,
while only the diagram with $Z^0$ boson gives a contribution to the SD
one. To convert coupling on quarks into couplings on nucleons we refer
to a standard set of parameters~\cite{Gasser,SMC} for nucleonic matrix
elements~\footnote{Note that the strange content here is slightly
smaller than the values implemented in other analyses,
see~\cite{Gondolo:2004sc,paololars} for details.}.

\FIGURE[t]{
\centerline{\epsfig{file=plot/logcros.eps,width=7.5cm} \quad
\epsfig{file=plot/lnspindep.eps,width=7.5cm}
}
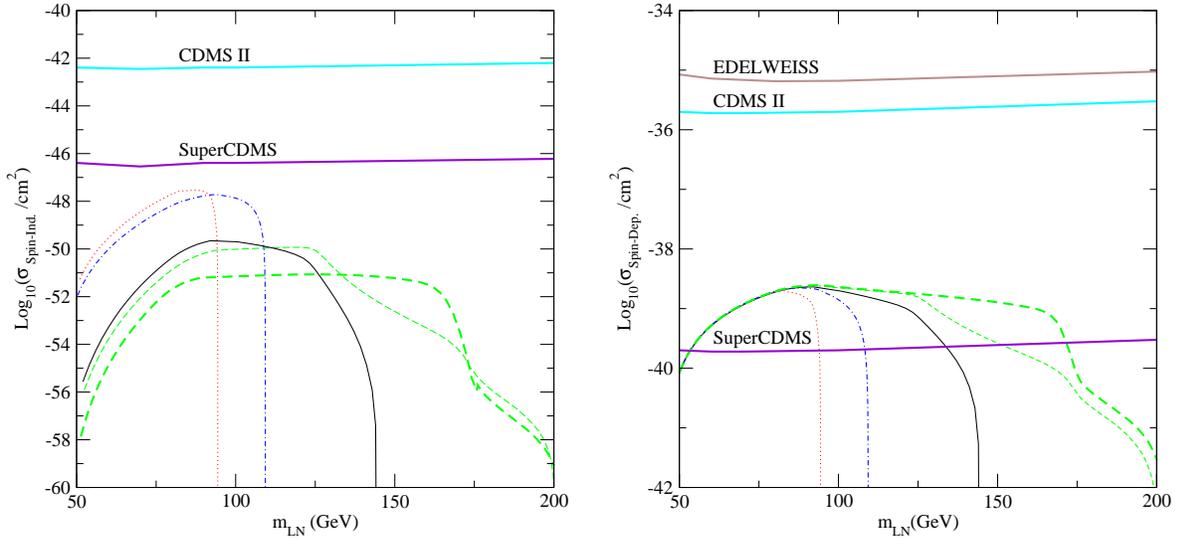
\caption{The spin-independent cross-section on a proton (left panel)
and the spin-dependent cross-section on a neutron (right panel) versus
the lightest neutralino mass, as compared to current exclusion curves
(CDMS~II) and the projected sensitivity of future detectors
(SuperCDMS). The models displayed are those at $\Omega_{LN} h^2
=0.113$ singled out in Fig.~\protect{\ref{fig:relic}}, with the same
sample choice of parameters and the same line-type (color) coding.
Note the mismatch in the vertical scale of the two plots.}
\label{fig:dd}}

In the left panel of Fig.~\ref{fig:dd} we plot predictions for the SI
neutralino-proton scattering cross-section $\sigma_{\chi P}^{\rm SI}$,
as a function of the LEF mass.  Models we display are those singled
out in Fig.~\ref{fig:relic} at $\Omega_{LN} h^2 =0.113$, with the same
sample parameters and coding therein.  For comparison we have shown
the exclusion curve from the null search by
CDMS~II~\cite{Akerib:2004fq} and the future expected sensitivity of
the SuperCDMS project in its one~ton
configuration~\cite{supercdms}~\footnote{The projected sensitivity of
other planned next-generation detectors of equal size, such as in the
setup of the XENON facility~\cite{Aprile:2002ef} is expected to be
comparable.}. As it can be seen predictions are orders of magnitude
below current sensitivities, as well as far below projected future
sensitivities, as one could have foreseen from the features of the
slice in the parameter space we have zoomed in. In fact as already
mentioned $\sigma_{\chi P}^{\rm SI}$ has one single contribution
mediated by a Higgs exchange, which is largely suppressed since the
$\neu1\neu1 H$ vertex scales with the gaugino-Higgsino mixing in
$\neu1$: in our guideline model the LN is always a very pure Higgsino,
its gaugino content going to zero in the limit $h_-\rightarrow 0$, and
it is of the order of few percent even for the largest $|h_-|$ reached
in Fig.~\ref{fig:relic} (the peaks in each of the displayed
curves). In all but one case we are considering a rather light Higgs,
$m_H = 150$~GeV. Since the cross-section scales with the inverse of
its forth power, taking $m_H$ equal to its current lower limit ($\sim
115$ GeV) one only gains a factor of about 3 in the
cross-section. Actually considering a heavier Higgs the prediction
gets rapidly further suppressed, as it can be seen by comparing the
case we plot with $m_H = 300$~GeV (thick dashed line) with the
corresponding one at $m_H = 150$~GeV (thin dashed line).

In the right panel of Fig.~\ref{fig:dd} we instead plot the
predictions for the SD neutralino-neutron scattering cross-section
$\sigma_{\chi N}^{\rm SD}$~\footnote{The search for SD couplings is
usually not listed as top priority for direct detection searches since
the lack of a coherent effect on the target nucleus dumps the
sensitivity with respect to the SI coupling, while in the MSSM frame
SI and SD terms usually have comparable strength. This is clearly not
the case in our setup.}. Here the picture looks more promising since
now the cross-section scales with the coupling of the LN with the
$Z$-boson, {\em i.e.}~the same effect setting, to some extent, the LN
relic density.  Again in each plot of $\sigma_{\chi N}^{\rm SD}$
versus mass there are maxima corresponding to the largest values of
$|h_-|$ along each isolevel curve. Such maxima are again well below
current sensitivities (the best exclusion curve being again set by the
CDMS~II result~\cite{cdmssd}), however perhaps within the reach of
future detectors.  The CDMS Collaboration is performing its DM
searches with natural $Ge$, which has a small component (around 8\%)
of the $^{73}Ge$ isotope, a target with an unpaired neutron from which
the limit on $\sigma_{\chi N}^{\rm SD}$ has been derived. We can
derive a rough projection for the gain in sensitivity on SD couplings
by simply scaling down the current exclusion curve of SuperCDMS
regarding the SI coupling in Ref.~\cite{supercdms} and shown in the
left panel~\footnote{An analogous sensitivity should be obtainable
with planned experiments using xenon.}. We find that a substantial
fraction of the models along the $\Omega_{LN}h^2$ isolevel curves in
our guideline framework, down to masses close to 50~GeV, will be
detectable by SuperCDMS or by an equivalent experiment.

\subsection{Neutrino telescopes}

\FIGURE[t]{
\centerline{\epsfig{file=plot/neurate.eps,width=8.1cm} 
}
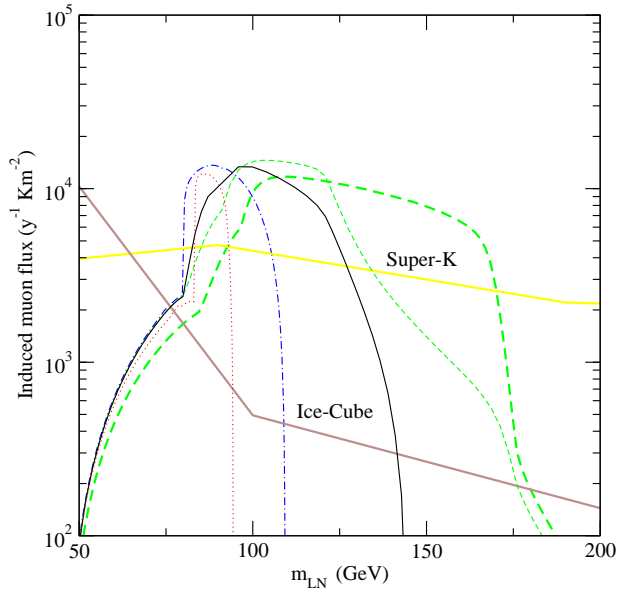
\caption{Muon flux induced by the neutrino flux from LN pair
annihilations in the center of the Sun versus LEF mass, and comparison
with the current best exclusion curve (SUPER-KAMIOKANDE) and the
projected sensitivity of future detectors (IceCube).  The models
displayed are those at $\Omega_{LN} h^2 =0.113$ singled out in
Fig.~\protect{\ref{fig:relic}}, with the same sample choice of
parameters and the same color (line-type) coding.}
\label{fig:neutel}}

The search for neutrinos produced by the annihilation of neutralinos
trapped in the core of the gravitational wells of the Sun or of the
Earth is a very promising indirect detection technique since it has a
very distinctive signature, and potentially induced fluxes may be
large.  In the present framework, since spin-independent
cross-sections are small, capture rates and fluxes from the Earth are
actually very low and will not be considered further. To estimate
neutrino fluxes from the Sun we implement the standard procedure
described in Refs.~\cite{Bergstrom:1998xh,joakimnt}, except for a more
careful treatment of neutralino capture
rates~\cite{sugrarates,Gondolo:2004sc}.  In Fig.~\ref{fig:neutel}, we
present results in terms of muon-induced fluxes, above the threshold
of 1~GeV, and compare them to the current best limits from the
SUPER-KAMIOKANDE Collaboration~\cite{Habig:2001ei} and with the future
projected sensitivity of the IceCube
experiment~\cite{icecube}~\footnote{The mismatch in the energy
threshold of IceCube and the threshold considered here has been taken
into account.}. The color coding on the $\Omega_{LN}$ isolevel curves
is the same as in Figs.~\ref{fig:relic} and~\ref{fig:dd}. Since the
capture rate in the Sun is driven by the SD neutralino-proton
coupling, and we have just verified that this can be fairly large, we
find that a large portion of LN models in our guideline scenario
sharply overshoots the currently best exclusion curve, and that there
are fair chances of detection with the improved sensitivity of
IceCube. The muon-induced fluxes sharply increase at the $W$
threshold, since LN annihilations at zero temperature into gauge boson
final states (which are a copious source of high energy neutrinos) is
not helicity suppressed as it happens for the $b\,\bar{b}$ final state
which dominates at lower masses.

Summarizing our result we find that, in the present guideline
framework, LN dark matter models slightly heavier than the $W$ mass
along the $\Omega_{LN} h^2$ isolevel curves corresponding to
$|h_-|\simgt 0.15$ are already excluded by current limits. This
conclusion essentially holds independently of the choice of the
parameter $m_H$ (compare the thick and thin dashed curves which differ
only in the value of $m_H$). In the future, with neutrino telescopes
it will be possible to test models at smaller values of $h_-$ in the
heavier mass branch, covering a region of parameter space larger than
the one expected from spin-dependent couplings in direct detection. On
the other hand since upcoming neutrino telescopes have a high energy
threshold no progress is foreseen in the light mass branch where,
instead, direct detection in the future will be more competitive.

\subsection{Halo rates}

Lightest neutralino pair annihilations in the Galactic halo may be a
significant source of cosmic-ray and gamma-ray fluxes. We will mainly
focus on the first ones and mention gamma-rays at the end of the
Section.

Charged particles injected in the Galaxy get trapped in the
interstellar magnetic fields building up an equilibrium population and
diffusing up to the solar system and the Earth where they can be
detected. Since there is no evidence of standard primary sources of
antimatter, and antimatter of secondary origin is scarce, searching
for antimatter from dark matter pair annihilations is a promising
technique to test the dark matter paradigm. We will consider the
induced antideuteron, antiproton and positron fluxes.

\FIGURE[t]{
\centerline{\epsfig{file=plot/chi2pb.eps,width=8.1cm}}
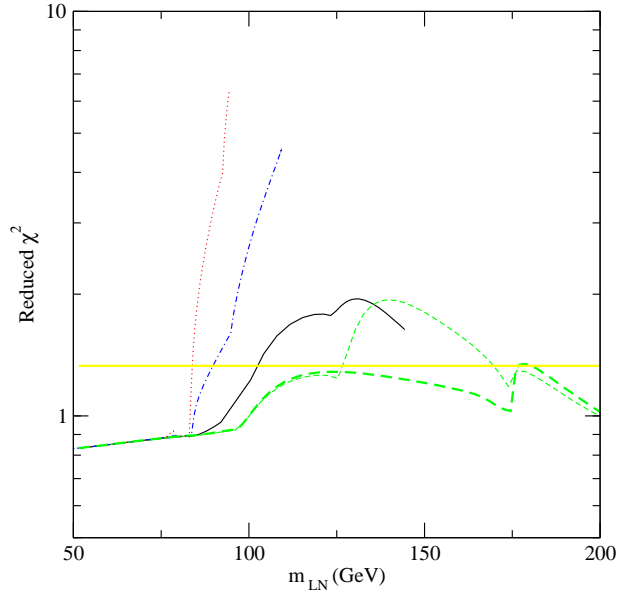
\caption{Reduced $\chi^2$ for the fit of presently available
antiproton flux data with a background plus neutralino signal, as a
function of lightest neutralino mass and within models singled out in
Fig.~\protect{\ref{fig:relic}}. The 3~$\sigma$ discrimination level is
shown as a horizontal line. The adiabatically contracted halo profile
has been assumed in this computation: considering the Burkert profile,
the signal becomes a small correction over the background and no model
can be discriminated.  }
\label{fig:chi2}}

Predictions involve several elements: the particle physics setup fixes
the pair annihilation cross-section $\sigma_{\rm ann}v$ and the
branching ratios for the various annihilation channels. For each of
them fragmentation and/or decay processes give rise to the stable
antimatter species, a step we include using tabulated results from the
\code{Pythia}~\cite{pythia} 6.154 Monte Carlo code as included in the
\ds package, except for $\bar{D}$ sources for which we have
implemented the prescription suggested in Ref.~\cite{dbar} to convert
from the $\bar{p}$-$\bar{n}$ yields.  To complete the estimate of the
strength of dark matter sources, one needs the number density of
neutralino pairs locally in space, {\em i.e.}~in terms of the dark
matter density profile $\rho$ and the dark matter particle mass
$m_{LN}$, as $1/2\,(\rho(\vec{x}\,) / {m_{LN}})^2$. The choice of the
halo profile is then crucial in the prediction of fluxes: we will
consider two possible setups ranging from the most favorable one for
dark matter detection to one of the least favorable ones.

For the first choice, which we will refer as the adiabatically
contracted model, we consider a model obtained by assuming that the
dark matter profile of the Milky Way, before gas cooling and the
formation of its luminous components, is described by the universal
profile found in Ref.~\cite{n03} resulting from N-body simulations of
hierarchical structure formation in a $\Lambda$CDM cosmology, with
mass $M_{vir} = 1.8 \times 10^{12}\msun$ and concentration parameter
$c_{vir} = 12$. The baryon infall is included assuming a smooth and
slow process, with a further enhancement of the dark matter density in
the central portion of the Galaxy (adiabatic contraction limit with no
redistribution of angular momentum between its
components~\cite{blumental}). The central portion of the profile
becomes as steep as $1/r^{1.5}$, but this singularity has been
conservatively cut off in its innest 1~pc, corresponding to a core
radius~\cite{ulliobh,milo} which is possibly induced by one of the
scenarios for the formation of the black hole sitting at the center of
the Galaxy~\cite{bhobservation}.

The second halo model we consider is defined by a profile with a large
core radius, a Burkert profile~\cite{burkert} with $M_{vir} = 1.3
\times 10^{12}\msun$ and $c_{vir} = 16$. We can think about this case
as the limit in which the profile is reshaped by a large
redistribution of angular momentum during the baryon infall, with the
inner density being sensibly reduced. Both profiles are assumed to be
spherical and to have analogous values of the local halo density,
$0.38$~GeV~cm$^{-3}$ and $0.34$~GeV~cm$^{-3}$, respectively. Hence
predictions in direct and indirect detection with neutrino telescopes
do not change appreciably and we have not anticipated this
discussion~\footnote{See Ref.~\cite{sugrarates} for further details on
the two halo models.}. The analysis could be more articulated
including effects {\em e.g.}~of substructures giving further
enhancements in the predictions. However we will not consider this
possibility here, and instead we will take a more conservative
approach.

The last step to make a prediction for the antimatter fluxes is to
model the propagation in the intergalactic magnetic fields and against
the solar wind within the solar system.  The propagation model adopted
for antiprotons and antideuterons has been developed in
Ref.~\cite{pbarpaper} and that for positrons in
Ref.~\cite{epluspaper}. Free parameters in both cases are set in
analogy to a setup which has been shown to reproduce fairly well the
ratios of primary to secondary cosmic ray nuclei~\cite{strmosk} with
the \code{Galprop}~\cite{galprop} propagation code. Solar modulation
is instead sketched with the analytical force-field
approximation~\cite{GleesonAxford}, with a modulation parameter as
appropriate at each phase in the solar cycle activity.

We first compare the prediction for the antiproton flux against a
compilation of data collected in the latest years. We consider results
with the BESS experiment that has measured with fairly good statistics
the antiproton flux in the energy range between 180 MeV and 4.2 GeV
during its flights in 1997, 1998, 1999 and 2000~\cite{bess}, and those
in the range between 3 and 50~GeV obtained by the CAPRICE experiment
during its 1998 flight~\cite{capricepbar}.  The expected component
from neutralino annihilations is added to the secondary component due
to cosmic-ray interactions, again estimated with the
\code{Galprop}~\cite{galprop} code under the same setup implemented
for the neutralino-induced component, that yields
$\chi^2=0.82$. Values of the reduced $\chi^2$ for the case signal plus
background are shown in Fig.~\ref{fig:chi2}, in the case of models
along the sample isolevel curves singled out in Fig.~\ref{fig:relic}
and for a distribution of dark matter particles according to the
adiabatically contracted profile. Some of the models give values of
the reduced $\chi^2$ as large as a few and are most probably excluded
by antiproton measurements. One should note however that the
3~$\sigma$ exclusion level at about 1.4 should not be considered a
strict bound since we have not taken into account uncertainties in the
propagation model and other steps in our prediction. At the same time
the limits we show are very sensitive to our halo choice: if the more
conservative Burkert halo is instead considered all signals get
suppressed by a factor of $\sim 100$, becoming a small correction with
respect to the background and leaving no chance of discrimination with
current data.

\FIGURE[t]{
\centerline{\epsfig{file=plot/adratiopb.eps,width=7.5cm} \quad
\epsfig{file=plot/adratioep.eps,width=7.5cm}}
\caption{Visibility parameter for future antiproton (left panel) and
positron (right panel) searches as compared to the detection
perspectives with the PAMELA instrument.  The models displayed are
those at $\Omega_{LN} h^2 =0.113$ singled out in
Fig.~\protect{\ref{fig:relic}}, with the same sample choice of
parameters and the same color (line-type) coding. Predictions are
shown in case of the adiabatically contracted halo profile.}
\label{fig:antimatter}}

Perspectives for the future are sketched in the left panel of
Fig.~\ref{fig:antimatter}. The quantity plotted on the vertical axis
is
\begin{equation}
I_{\Phi} \equiv \int_{E_{min}}^{E_{max}} dE \,
\frac{\left[\Phi_s(E)\right]^2}{\Phi_b(E)}\,,\label{eq:visibility}
\end{equation}
where $\Phi_s(E)$ and $\Phi_b(E)$ are the signal and background
fluxes, respectively, and the integral extends over the whole interval
in which the ratio is non negligible.  It gives the continuum limit of
a $\chi^2$-like variable in the regime in which the signal is a small
correction to some known background, see Ref.~\cite{stefanopiero} for
details.  In Fig.~\ref{fig:antimatter} the horizontal line gives, in
this same variable, the discrimination level which will be reached by
the PAMELA experiment~\cite{pamela} in three years of data taking
(which should start in early 2006). The predictions are for the
adiabatically contracted profile and they indicate that in such setup
all models with the LN heavier than the $W$ gauge boson, even those
with extremely small $|h_-|$, would be tested. This signal, as all
halo signals, scales with the total annihilation rate at zero
temperature which for gauge boson final states is unsuppressed and
little related to the coupling of the LN with the $Z^0$, unlike in the
case of lower masses and fermion final states. Again we must stress
that this conclusion heavily relies on which halo profile is chosen:
if the Burkert profile is implemented, predictions for the parameter
$I_{\Phi}$ are shifted down over two orders of magnitude, and no model
would be tested even in the future within such a setup.

In the right panel of Fig.~\ref{fig:antimatter} we show the analogous
picture for the positron fluxes. The range of models which will be
testable in the future, for the adiabatically contracted profile, is
in this case slightly smaller than in the antiproton case. Limits from
current data do not allow any model discrimination even with this halo
model, and hence the analogue of Fig.~\ref{fig:chi2} is not shown.

\FIGURE[t]{
\centerline{\epsfig{file=plot/adgapheart.eps,width=7.5cm} \quad
\epsfig{file=plot/gaps.eps,width=7.5cm}
}
\caption{Visibility ratio for future antideuteron searches with the
GAPS instrument (all models above the horizontal line are detectable),
in case of a mission with a satellite on an earth orbit. The
distribution of neutralinos in the halo has been assumed according to
the adiabatically contracted model (left panel) or the Burkert profile
(right panel).  The models displayed are those at $\Omega_{LN} h^2
=0.113$ singled out in Fig.~\protect{\ref{fig:relic}}, with the same
sample choice of parameters and the same color (line-type)
coding. Values of the visibility ratio shift up by a factor of about 3
if one considers the same instrument placed on a deep space probe. }
\label{fig:dbar}}

At present there are no data on the antideuteron cosmic ray flux and
indeed, if one constrains oneself to the low energy window, the
secondary background flux is expected to be negligible~\cite{dbar}, so
that even the detection of one single event could be used to claim the
presence of an exotic primary source. To address the detection
prospects for the future we consider the gaseous antiparticle
spectrometer (GAPS)~\cite{Mori}, which will have the capability of
searching for antideuterons in the energy interval 0.1-0.4 GeV per
nucleon, with an estimated sensitivity level of
$2.6\times10^{-9}\textrm{m}^{-2}\textrm{sr}^{-1}\textrm{GeV}^{-1}
\textrm{s}^{-1}$, and that has been proposed as an instrument to be
placed in a satellite on an earth orbit or on a deep space
probe. Visibility ratios, {\em i.e.}~the ratio of the predicted flux
over the sensitivity, are shown in Fig.~\ref{fig:dbar} for the same
sample of models considered so far.  In the left panel the
adiabatically contracted halo profile is considered, while in the
right panel there are predictions with the Burkert profile. Note that
through this detection method, not only all models are essentially
found to be detectable when one considers the most favorable halo
profile, but also by taking the very conservative Burkert profile one
finds that large portions of parameter space, including part of the
$h_- =0$ regime, are testable. Furthermore the results displayed hold
for an instrument mounted in a satellite on an earth orbit, in a
location within the solar system at which a significant portion of low
energy antideuterons is diverted by the solar wind~\footnote{Solar
modulation are implemented as appropriate for a period close to solar
maximum.}.  Were the deep space experiment realized, all visibility
ratios would shift up by a factor of 3 or so, making most models
testable even under pessimistic assumptions for the halo profile.  It
then emerges clearly that the search for cosmic-ray antideuterons is
one of the most solid and competitive ways of testing our scenario in
the future.

\FIGURE[t]{ \centerline{\epsfig{file=plot/2gamma.eps,width=7.5cm}
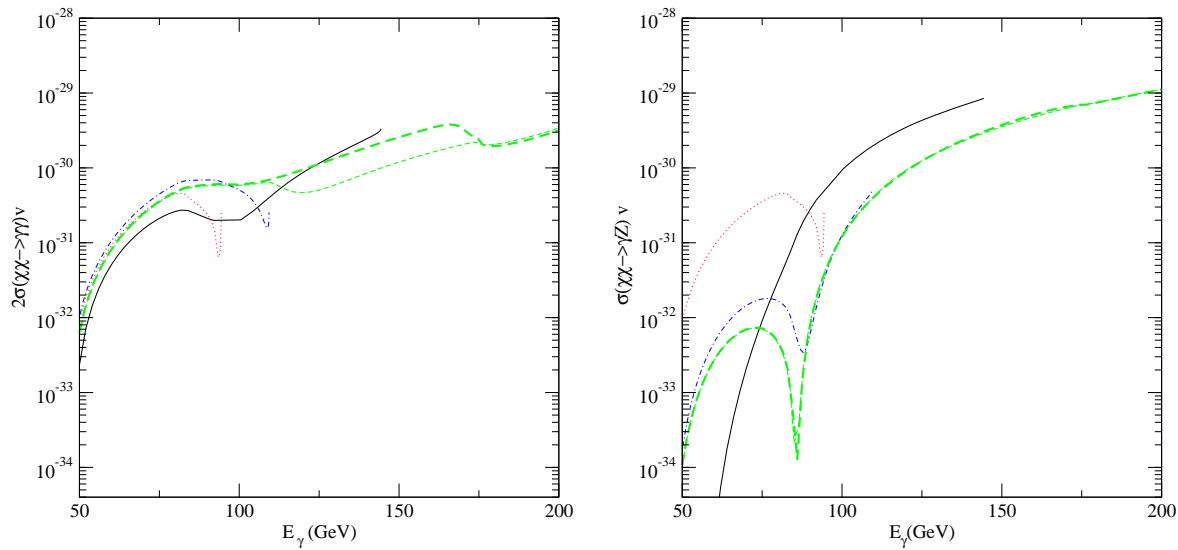
\quad \epsfig{file=plot/zgamma.eps,width=7.5cm}\caption{The
annihilation rate into two photons (left-panel) and into a $Z^0$ plus
a photon (right-panel) times the number of photons in the final state
versus the energy of the monochromatic photon.  The models displayed
are those at $\Omega_{LN} h^2 =0.113$ singled out in
Fig.~\protect{\ref{fig:relic}}, with the same sample choice of
parameters and the same color (line-type) coding. } }
\label{fig:line}
}

The fragmentation of final states from neutralino annihilations gives
as well neutral pions, which mainly decay into two photons. Gamma-rays
obtained in this channel can be rather copious but unfortunately they
have a weak spectral signature. The prediction for fluxes and the
possibility of angular discrimination of the signal are very tightly
correlated to details in the distribution of dark matter particles in
the very central region of the Galaxy, which is essentially
unknown. Actually it is more interesting to check whether the process
of prompt emission of photons through loop induced annihilation
processes~\cite{lines}, which gives a monochromatic gamma-ray flux
with no conceivable standard astrophysical counterpart, is effective
or not. In Fig.~\ref{fig:line} we plot annihilation rates times number
of monochromatic photons in the final state versus the energy of the
photon, for two such possible final states in the case of neutralino
annihilations, {\em i.e.}~the $\gamma\gamma$ and $Z^0\gamma$
processes.  Values for the rates are at the level of the largest value
one can obtain for thermal relic neutralinos in the
MSSM~\cite{BUB}. Hence detection prospects in this channel should be
comparable to MSSM cases, {\em i.e.}~feasible in some configurations
but not as promising as some of the techniques we described so far.

\section{Towards a more generic model within the framework}
\label{sec:towards}

All the previous discussion relies on the simplifying assumptions of
Bino decoupling and projection along the direction $M_2=-|\mu|$. This
automatically drove the LN to be in a rather pure Higssino state. The
small gaugino--Higgsino mixing induced negligibly small
spin-independent scattering cross-sections, while the large couplings
to gauge bosons gave fairly large indirect detection rates and small,
but larger than usual and possibly detectable, spin-dependent
couplings. We now wish to check how the picture changes when moving in
other directions of our parameter space.  We will consider a case with
larger gaugino-Higgsino mixing. We relax the relation $M_2=-|\mu|$ and
let again the Bino-like neutralino to be light and coupled to the
other particles.  To deal with a reasonable number of parameters we
restrict ourselves to the subspace defined by: {\sl a)} $\mu$ real,
since a non-vanishing phase is only affecting the mass spectrum; {\sl
b)} $M_1=\alpha M_2$; {\sl c)} $h_+=1.5$, the minimum value needed to
obtain baryogenesis~\cite{Carenaetal}; {\sl d)} $h_-=-0.125$, a limit
in which the Higgsino coupling to the $Z$ boson is suppressed but
non-negligible; {\sl e)} $m_H=150$~GeV, {\em i.e.}~close to the
presently preferred value from electroweak precision
measurements~\cite{PDG}; {\sl f)} $h'_1=h'_2=0.25$ in order to get a
$T$ parameter value in agreement with electroweak precision
measurements~\footnote{For $m_H=150$ GeV a fit to the precision
electroweak data has been done by the LEP electroweak working group
yielding~\cite{Carenaetal}
$$S=0.04\pm0.10,\quad T=0.12\pm0.10$$ with an 85\% correlation between
the two parameters. We have checked that for values of $h_i$ and
$h'_i$ of $\mathcal{O}(1)$ the contribution to the $T$ parameter is
$\mathcal{O}(1)$ and the corresponding models are thus excluded by
precision data for any value of the Higgs mass. However for the
previous values of $h_i$ and $h'_i$ we find $T\simlt 0.2-0.3$
depending on the values of the masses $M_{1,2}$ and $\mu$, that can be
accommodated into the present electroweak bounds depending on the value
of the Higgs mass.}. We are thus left with three free parameters:
$\mu,M_2,\alpha$.

\FIGURE[t]{
\centerline{\epsfig{file=plot/M2vsmu.eps,width=8.1cm}}
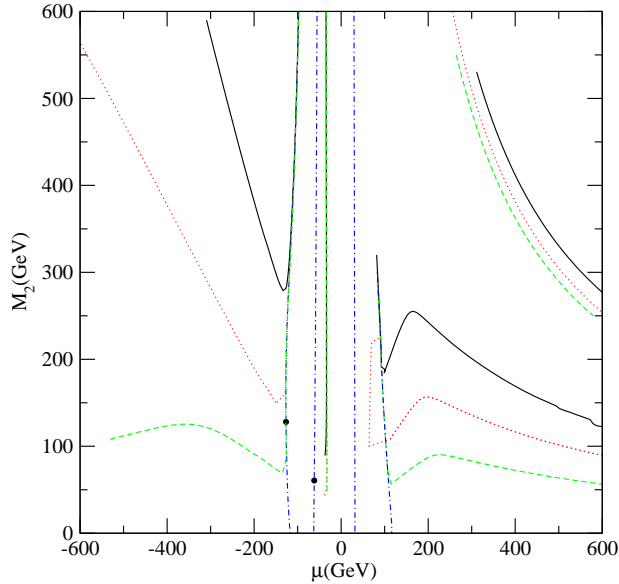
\caption{Sample isolevel curves of the lightest neutralino relic
abundance at the WMAP preferred value $\Omega_{LN} h^2 = 0.113$ and in
agreement with the $Z$ boson width measurement. From top to bottom,
the extra free parameters are set equal to: $M_1=M_2/2$ black (solid)
line, $M_1=M_2$ red (dotted) line, $M_1=2M_2$ green (dashed) line,
$M_1=1$ TeV blue (dash-dotted) line, the Bino decoupling limit. Black
dots show models in the guideline-case.\label{fig:M2vsmu}}}

In Fig.~\ref{fig:M2vsmu} we show isolevel curves of the lightest
neutralino relic abundance in the plane $(\mu;\,M_2)$ with the ratio
$\alpha=M_1/M_2$ equal to 0.5, 1 and 2, along with the Bino-decoupling
limit at large $M_1$ ($\alpha\to\infty$). The LN thermal relic density
is computed using the same procedure described in
Section~\ref{sec:guide}. Each of the four cases above considered
corresponds in turn to four branches of isolevel curves. There are two
regimes depending on whether $M_2$ and $\mu$ have the same or opposite
signs and, for each one, the pair of isolevel curves delimits the
region where the relic density is exceeding the cosmologically
preferred value.

Let us focus {\em e.g.}~on the case $M_1=M_2/2$, {\em i.e.}~the solid
lines in the plot. As it can be checked from the general results of
Section~\ref{sec:fram} at $\mu>0$, in the top-right corner, we find
the branch corresponding to large Higgsino-Bino mixing, with both
neutralino and chargino masses of the order of 40~GeV. Since the lower
bound on the chargino mass is $\sim 104$~GeV~\cite{PDG} these models
are ruled out. The second branch starts at large values of $M_1$,
where the neutralino is Higgsino-like and slightly heavier than the
$W$ boson. It extends down to smaller and smaller values of $M_1$ on a
nearly vertical path along which the Higgsino purity monotonically
decreases.  Then the branch bends along a quasi-horizontal path in
which the neutralino turns into a pure gaugino, with a predominant
Bino component; in this case the relic abundance is settled by
annihilation in $W$ bosons and co-annihilation with NLN.  At $\mu<0$,
for large $M_1$, as in the guideline model we discussed in previous
sections, a given $h_-$ selects two models with a Higgsino-like LN and
equal relic density: one with a mass smaller than the $W$ mass,
annihilating into fermions, and another one with a larger mass, mainly
annihilating into gauge bosons. The heavier branch starts at high
values of $M_1$, decreases monotonically and reaches a minimum value
of $M_1$. Now the LN is an almost pure Bino and hence the coupling
with the $Z$ boson is suppressed: the annihilation channels in
fermions and $Z$ bosons are less effective. The curve rises again to
high values of $M_1$ following an oblique path, along which the relic
abundance is essentially settled by co-annihilation with NLN and LC. 

In the case $M_1=2M_2$ the behavior is different: the isolevel curves
follow quasi-horizontal paths in which the relic abundance is fixed
through its Wino component and co-annihilation with NLN.  Finally, in
the Bino decoupling case, for both signs of $\mu$, cosmologically
interesting models are located in the region where the LN is
Higgsino-like.

\subsection{Direct detection}

\FIGURE[t]{
\centerline{\epsfig{file=plot/loghprime025spinindep.eps,width=7.5cm} \quad
\epsfig{file=plot/lnm1alpm2spindip.eps,width=7.5cm}
}
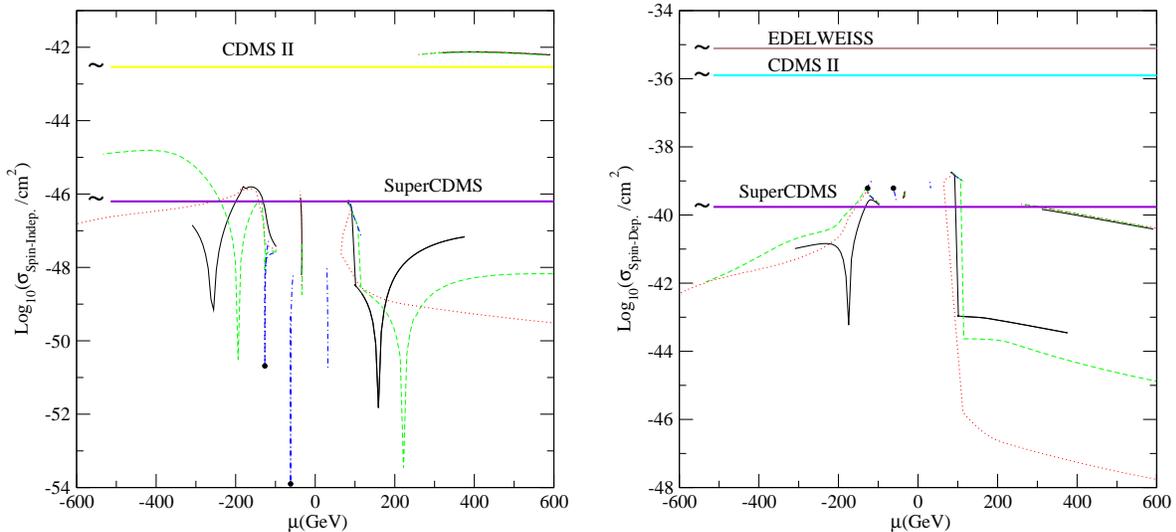
\caption{The spin-independent cross-section on a proton (left panel)
and the spin-dependent cross-section on a neutron (right panel) versus
the parameter $\mu$, and comparison with the approximate excluded
value in the relative LN mass range and the sensitivity level of
future detectors. The models displayed are those at $\Omega_{LN} h^2
=0.113$ singled out in Fig.~\protect{\ref{fig:M2vsmu}}, with the same
sample choice of parameters and the same color (line-type) coding.
Black dots show models in the guideline-case.  Note the mismatch in
the vertical scale of the two plots.}
\label{fig:m1alpm2dd}
}

Following the discussion of the guideline model we now want to
investigate the detection prospects of the LN in this extended
framework. We start again with direct detection and present
predictions in terms of elastic scattering cross-sections off of a
single nucleon, separating SI interactions (given by a Higgs boson
exchange) from SD interactions (induced by a Z boson exchange).  

In the left panel of Fig.~\ref{fig:m1alpm2dd} we present the
prediction for the SI neutralino proton scattering cross-section
versus the parameter $\mu$, for models along the relic density
isolevel curves singled out in Fig.~\protect{\ref{fig:M2vsmu}}. One
should notice that now the correspondence between $\mu$ and LN mass is
lost, being the LN mass of order 100-300~GeV for almost all displayed
models. Still since we are not referring to a single mass but to a
(small) mass range the comparison with experimental limits and future
expected sensitivities is approximate and indicative. As expected the
SI scattering cross-sections for LN models with large gaugino-Higgsino
mixing can be much enhanced compared to the corresponding ones in the
guideline model. We find that models at large positive $M_2,\,\mu$ are
actually already excluded by present CDMS II data, while there is a
fair fraction of models with SI cross-sections exceeding the
sensitivity level of Super-CDMS. Note also that along branches at
which the LN is mostly gaugino-like there can be, depending on the
value of $\mu$, an accidental cancellation in the LN coupling to the
Higgs driving sharp falls in the SI cross-section.

In the right panel of Fig.~\ref{fig:m1alpm2dd} we plot the predictions
for the SD neutralino-neutron scattering cross-section. In this case
the relevant quantity is the coupling strength to the $Z$ boson,
involving only Higgsino states and projecting out the mixing between
the two. This tends to always be smaller than in the guideline case,
see Eq.~(\ref{eq:zcoup}). Nevertheless a fraction of the parameter
space is within the projected sensitivity of SuperCDMS.

\FIGURE[t]{
\centerline{\epsfig{file=plot/m1alpm2neurate.eps,width=8.1cm}}
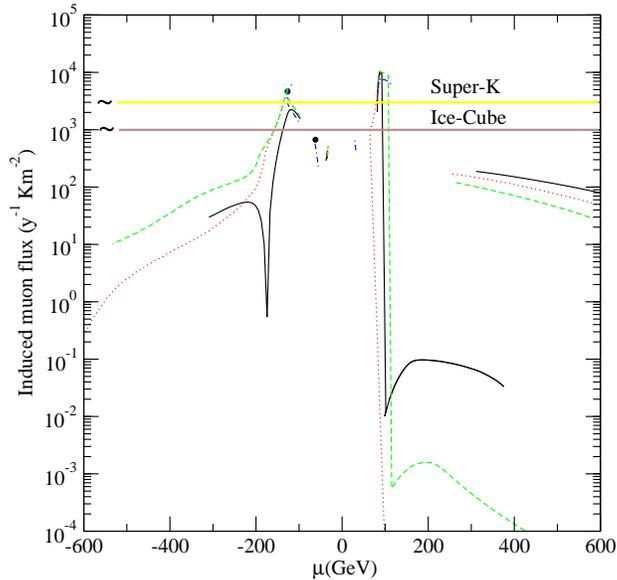
\caption{Muon induced flux due to annihilations of lightest neutralino
pairs in the center of the Sun versus the parameter $\mu$, as compared
to the level excluded by SUPER-KAMIOKANDE assuming an average mass
value and the projected sensitivity of IceCube for the same mass
value.  The models displayed are those at $\Omega_{LN} h^2 =0.113$
singled out in Fig.~\protect{\ref{fig:M2vsmu}}, with the same sample
choice of parameters and the same color (line-type) coding. Black dots
show models in the guideline-case.}
\label{fig:m1m2neutel}}

\subsection{Indirect detection}

The prospects to detect the LN with neutrino telescopes is discussed
in Fig.~\ref{fig:m1m2neutel}, where we plot the induced muon flux
versus $\mu$. Even in this plot the experimental limits are just
indicative because a direct link between the LN mass and $\mu$ is
missing.  Since the capture rate in the Sun scales with the neutralino
spin-dependent cross-section, we find the induced neutrino flux to be
very small for models with large $|\mu|$.  As for spin-dependent
couplings, the perspectives of detection are worse than in the
guideline model, with few cases above the projected sensitivity of
IceCube.

Following the same approach discussed in Section~\ref{sec:guide}, we
compute the expected positron and antiproton fluxes originated by
neutralino annihilation in the Galactic halo. In
Fig.~\protect{\ref{fig:m1m2antimatter}} we show the results in the
case of the adiabatically contracted halo. We plot the visibility
parameter we defined in E.~(\protect{\ref{eq:visibility}}) versus
$\mu$. Using this detection method the most promising models are those
with a large annihilation cross-section in $W$ bosons, {\em i.e.}~the
upper branch in the $\mu<0$ region of
Fig.~\protect{\ref{fig:M2vsmu}}. This also explains the behavior of
the signal in the other regimes. In fact, the lower isolevel branch at
$\mu<0$ corresponds to values of the LN mass at which the only open
annihilation channels are those into fermions.  In the case $\mu>0$
the situation is slightly different because, considering the lower
branch, starting from high values of $M_2$ and moving along the
isolevel curve the cross-section in $W$ bosons decreases since the LC
is becoming heavier. Along the horizontal paths we have the opposite
behavior: in fact the LC is becoming lighter.  This explains the
behavior of the curves in Fig.~\protect{\ref{fig:m1m2antimatter}}.
Only a fraction of the models is detectable with the upcoming PAMELA
experiment. Predictions are again relying on the distribution of dark
matter in the halo and, in case a Burkert profile is considered, all
fluxes drop off by a factor of about ten, driving all estimates for
the visibility parameter below the sensitivity of PAMELA.

\FIGURE[t]{
\centerline{\epsfig{file=plot/m1alpm2adpb.eps,width=7.5cm} \quad
\epsfig{file=plot/m1alpm2adeb.eps,width=7.5cm}
}
\caption{Visibility ratio for future antiproton (left panel) and
positron (right panel) searches as compared to the detection
perspectives with the PAMELA instrument.  The models displayed are
those at $\Omega_{LN} h^2 =0.113$ singled out in
Fig.~\protect{\ref{fig:M2vsmu}}, with the same sample choice of
parameters and the same color (line-type) coding. Black dots show
models in the guideline-case. }
\label{fig:m1m2antimatter}}

The trends we sketched also hold for antideuteron searches. In
Fig.~\protect{\ref{fig:m1m2dbar}} we plot the GAPS visibility ratio
versus $\mu$.  The shape of the lines is very similar to those in
Fig.~\protect{\ref{fig:m1m2antimatter}}.  In the left panel we
consider the configuration with the instrument on a satellite around
the earth while in the right panel we consider the case of a probe in
the deep space. As in the previous discussion the prospects for this
kind of searches are more favorable in case of annihilation dominated
by gauge boson final states.

The last case we have studied is the monochromatic gamma ray
production by neutralino annihilation for two photons and $Z$ boson
plus photon final states, as shown in
Fig.~\protect{\ref{fig:m1m2gamma}}. As discussed in
Section~\ref{sec:guide} these are one loop-processes via chargino or
SM fermions.  The most promising branches are in the $\mu<0$ region,
since the LC running in the loop is quasi-degenerate in mass with the
LN. This branch lies in an interesting energy range for upcoming
gamma-rays detectors like GLAST~\cite{glast}.

\FIGURE[t]{
\centerline{\epsfig{file=plot/m1alpm2adgapsearth.eps,width=7.5cm} \quad
\epsfig{file=plot/m1alpm2adgapsdeep.eps,width=7.5cm}
}
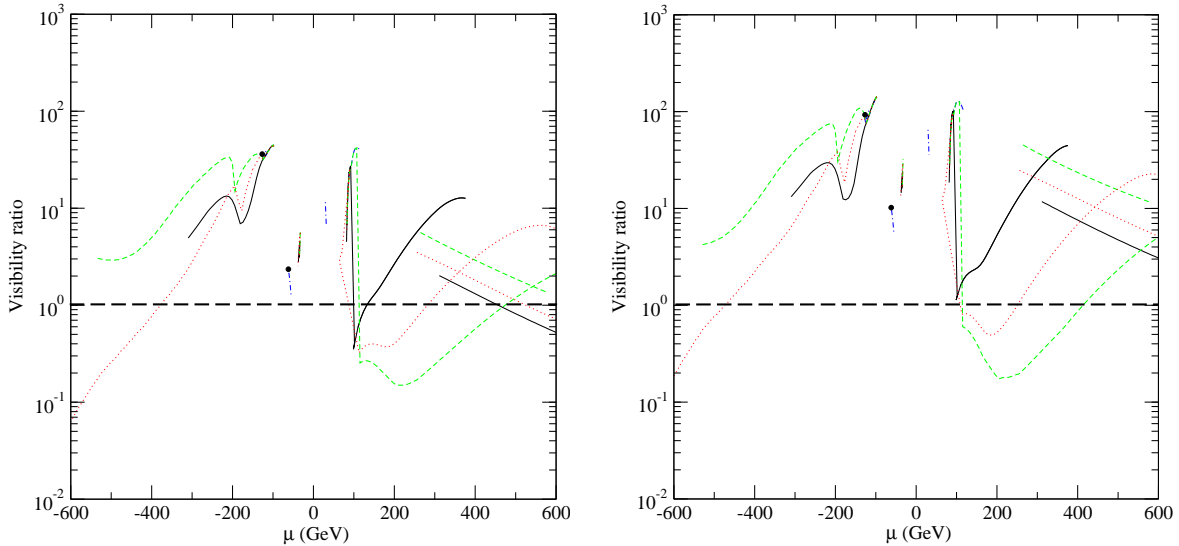
\caption{Visibility ratio for future antideuteron searches with the
GAPS instrument (all models above the horizontal line are detectable)
in the case of a mission with a satellite on an earth orbit (left
panel) and that of a deep space probe (right panel).  The models
displayed are those at $\Omega_{LN} h^2 =0.113$ singled out in
Fig.~\protect{\ref{fig:M2vsmu}}, with the same sample choice of
parameters and the same color (line-type) coding. Black dots show
models in the guideline-case. }
\label{fig:m1m2dbar}}
\FIGURE[t]{\vspace{0.5cm}
\centerline{\epsfig{file=plot/hprime0252gamma.eps,width=7.5cm} \quad
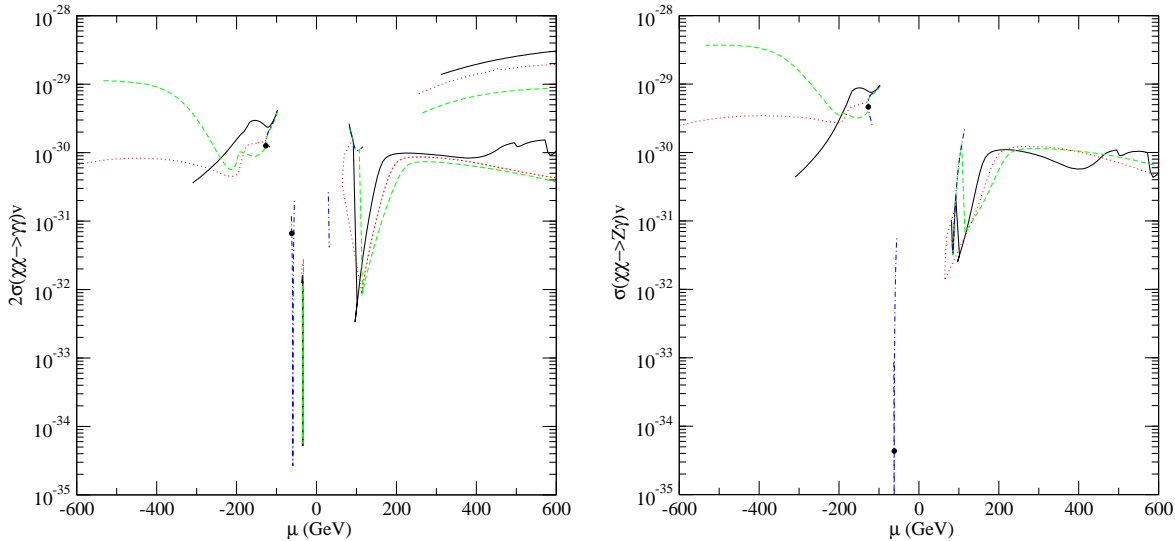
\epsfig{file=plot/hprime025zgamma.eps,width=7.5cm}\caption{The
annihilation cross-section in two photons (left-panel) and Z plus
photon (right-panel). The models displayed are those at $\Omega_{LN}
h^2 =0.113$ singled out in Fig.~\protect{\ref{fig:M2vsmu}} , with the
same sample choice of parameters and the same color (line-type)
coding. Black dots show models in the
guideline-case. }\label{fig:m1m2gamma} }}

\section{Conclusions}
\label{sec:conclusion}

Models with extra fermions and large Yukawas were introduced in the
context of electroweak baryogenesis. In this paper we have focused on
their implications on the dark matter problem.  In fact the particle
content of such models allows for the presence of a weakly interacting
massive particle that could be a good dark matter candidate. The
general setup of the model resembles a split supersymmetry scenario
where the supersymmetric relation between the Yukawas and gauge
couplings is relaxed. Bounds coming from the thermal relic abundance
select a light spectrum. In particular the chargino mass is typically
of the order of $200$ GeV, a favorable case for detection of physics
beyond the standard model at upcoming colliders.

We have separated our discussion into two parts. In the first one we
studied, as a reference model, the setup with a reduced number of
parameters introduced in Ref.~\cite{Carenaetal} to strengthen the
electroweak phase transition and achieve baryogenesis: it describes a
framework with Bino decoupling and the lightest neutralino being a
very pure Higgsino state. We have foliated the parameter space
retaining all the models with $\Omega_{LN} h^2$ in agreement with both
the WMAP determination and the $Z$ boson width measurement at LEP. We
computed the rates for direct and indirect detection. Due to the low
gaugino-Higgsino mixing, spin-independent elastic scattering
cross-sections are very small and they are not within the projected
sensitivity of planned detectors. On the other hand the spin-dependent
cross-sections, even if they are out of reach of the present
experiments, may be detected by future experiments. Indirect detection
techniques look more promising. In fact using data on the neutrino
flux from the center of the Sun we are able to rule out part of the
parameter space. The induced antimatter components in cosmic rays give
complementary and promising signals. In particular we predict that for
most models the antideuteron flux will be detectable with GAPS,
regardless on the assumptions on the dark matter distribution in the
Galaxy.

In the second part of this work we have extended our framework
allowing for a non-vanishing Bino mixing. In such a case the
neutralino mass matrix has more free parameters than the MSSM and,
with a suitable choice of them, one can induce a Bino-Wino mixing, a
configuration which is never realized in the MSSM.  In the case of a
large gaugino-Higgsino mixing, we obtain an increase of the
spin-independent cross-section and we can actually rule out models
with the largest mixing, with a general improvement of prospects for
this kind of searches at future experiments. Indirect detection
becomes less promising than in the guideline framework but using
neutrino data we can still rule out part of the parameter space. The
search for antimatter is promising for models with a lightest
neutralino mass above the $W$ threshold. Hence as bottom line to this
analysis and contrary to the standard lore, indirect detection
techniques generically seem the more promising strategies to detect
dark matter.

\section*{Acknowledgments}
This work was supported in part by CICYT, Spain, under contracts
FPA2004-02012, FPA2002-00748 and FPA2005-02211, and in part by
INFN-CICYT under contract INFN04-02.


\begin{thebibliography}{99}

\bibitem{ewb}  For reviews, see, e.g.:
A.G.~Cohen, D.B.~Kaplan and A.E.~Nelson,
 Ann. Rev. Nucl. Part. Sci. {\bf 43} (1993) 27;
M.~Quir\'os, \HPA{67}{1994}{451};
V.A.~Rubakov and M.E.~Shaposhnikov,  Phys. Usp. {\bf 39} (1996) 461;
M.~Carena and C.E.M.~Wagner, hep-ph/9704347;
A.~Riotto and M.~Trodden,  Ann. Rev. Nucl. Part. Sci. {\bf 49}
(1999) 35; M.~Quir\'os, hep-ph/9901312;
M.~Quir\'os and M.~Seco,  Nucl. Phys. {\bf B Proc. Suppl. 81} (2000) 63,
hep-ph/9703274.
%

\bibitem{wimps}
Among early references, see, e.g.:
B.~W.~Lee and S.~Weinberg, Phys. Rev. Lett. {\bf 39} (1977) 165;
J.~E.~Gunn et al., Astrophys. J. {\bf 223} (1978) 1015;
G.~Steigman et al., Astron. J. {\bf 83} (1978) 1050;
J.~Ellis et al., nucl. Phys. B {\bf 238} (1984) 453.

%
\bibitem{CQW} M.~Carena, M.~Quir{\'o}s and C.E.M.~Wagner,
\PLB{380}{1996}{81}.

\bibitem{Carenaetal}
  M.~Carena, A.~Megevand, M.~Quiros and C.~E.~M.~Wagner,
  Nucl. Phys. B {\bf 716} (2005) 319.

\bibitem{split}
N.~Arkani-Hamed and S.~Dimopoulos,
arXiv:hep-th/0405159;
G.~F.~Giudice and A.~Romanino,
arXiv:hep-ph/0406088.
%

\bibitem{Gondolo:2004sc}
P.~Gondolo, J.~Edsjo, P.~Ullio, L.~Bergstrom, M.~Schelke and E.~A.~Baltz,
JCAP {\bf 0407}, 008 (2004)
[arXiv:astro-ph/0406204].

\bibitem{Spergel:2003cb}
D.~N.~Spergel {\it et al.}  [WMAP Collaboration],
Astrophys.\ J.\ Suppl.\  {\bf 148} (2003) 175
[arXiv:astro-ph/0302209].
\bibitem{PDG}
  S.~Eidelman {\it et al.}  [Particle Data Group],
  Phys.\ Lett.\ B {\bf 592} (2004) 1.
\bibitem{jkg} 
  G.~Jungman, M.~Kamionkowski and K.~Griest,
  Phys. Rep. {\bf 267} (1996) 195.

\bibitem{Bergstrom:1998xh}
L.~Bergstrom, J.~Edsjo and P.~Gondolo,
Phys.\ Rev.\ D {\bf 58} (1998) 103519
[arXiv:hep-ph/9806293].

\bibitem{dirdet}
M.~W.~Goodman and E.~Witten, Phys. Rev.\ D {\bf 31} (1986) {3059};
I. Wasserman, Phys. Rev.\ D {\bf 33} (1986) {2071}.

\bibitem{Gasser} 
  J. Gasser, H. Leutwyler and M.E. Sainio,
  Phys. Lett. {\bf B253} (1991) 252.

\bibitem{SMC} 
  D. Adams et al, Phys. Lett. {\bf B357} (1995) 248.

\bibitem{paololars}
 L. Bergstrom and P. Gondolo, Astropart. Phys. {\bf 5} (1996) 263.

\bibitem{Akerib:2004fq}
D.~S.~Akerib {\it et al.}  [CDMS Collaboration],
arXiv:astro-ph/0405033.

\bibitem{supercdms}
P.~L.~Brink et al., The SuperCDMS Collaboration, astro-ph/0503583.

\bibitem{Aprile:2002ef}
E.~Aprile {\it et al.},
arXiv:astro-ph/0207670.

\bibitem{cdmssd}
L. Baudis, astro-ph/0503549.

\bibitem{joakimnt}
 L.~Bergstr\"om, J.~Edsj\"o and P.~Gondolo,
 Phys.\ Rev.\  {\bf D58} (1998) 103519.

\bibitem{sugrarates}
J.~Edsjo, M.~Schelke and P.~Ullio,
JCAP {\bf 0409}, 004 (2004) [arXiv:astro-ph/0405414].

\bibitem{Habig:2001ei}
A.~Habig  [Super-Kamiokande Collaboration],
arXiv:hep-ex/0106024.

\bibitem{icecube}
 J.~Edsj\"o, internal Amanda/IceCube report, 2000.


\bibitem{pythia}
T.~Sj\"{o}strand,
Comput.\ Phys.\ Commun.\  {\bf 82} (1994) 74. ;
T.~Sj\"{o}strand, {\em PYTHIA 5.7 and JETSET 7.4. Physics and Manual},
CERN-TH.7112/93, \hepph{9508391} (revised version).

\bibitem{dbar} 
 F. Donato, N. Fornengo and P. Salati, Phys. Rev. {\bf D62} (2000) 043003.

\bibitem{n03}
 J.F.~Navarro et al., MNRAS (2004) in press, astro-ph/0311231.

\bibitem{blumental}
  G.R.~Blumental, S.M. Faber, R.~Flores and J.R.~Primack,  
  Astrophys. J. {\bf 301} (1986) 27.

\bibitem{ulliobh}
  P. Ullio, H.S. Zhao and M. Kamionkowski, Phys. Rev. {\bf D 64}  
  (2001) 043504.

\bibitem{milo}
  D. Merritt, M. Milosavljevic, L. Verde and R. Jimenez,
  Phys. Rev. Lett. {\bf 88} (2002) 191301.

\bibitem{bhobservation}
 A.M.~Ghez et al, astro-ph/0306130.

\bibitem{burkert}
  A.~Burkert, Astrophys. J. {\bf 447} (1995) L25.

\bibitem{pbarpaper} 
 L. Bergstr\"om, J. Edsj\"o and P. Ullio, Astrophys. J. {\bf 526}
 (1999) 215.

\bibitem{epluspaper}
 E.A. Baltz, J. Edsjo, Phys. Rev. {\bf D59} (1999) 023511.

\bibitem{strmosk}
 I.V.~Moskalenko, A.W.~Strong, J.F.~Ormes and M.S.~Potgieter, 
 Astrophys. J. {\bf 565} (2002) 280.

\bibitem{galprop} 
 Galprop numerical package, 
 http://www.mpe.mpg.de/\~{ }aws/propagate.html


\bibitem{GleesonAxford} L.J.~Gleeson and
 W.I. Axford, Astrophys. J. {\bf 149} (1967) L115.


\bibitem{bess}
S.~Orito et al., Phys. Rev. Lett. {\bf 84} (2000) 1078;
Y.~Asaoka et al., Phys. Rev. Lett. {\bf 88} (2002) 05110.
\bibitem{capricepbar}
Boezio et al., Astrophys. J. {\bf 561} (2001) 787.

\bibitem{pamela}
     O.~Adriani et al. ({\sc Pamela} Collaboration),
     Proc. of the 26th ICRC, Salt Lake City, 1999, OG.4.2.04.

\bibitem{stefanopiero}
  S.~Profumo and P.~Ullio,
  JCAP {\bf 0407} (2004) 006
  [arXiv:hep-ph/0406018].


\bibitem{Mori}
  K.~Mori, C.~J.~Hailey, E.~A.~Baltz, W.~W.~Craig, M.~Kamionkowski, W.~T.~Serber and P.~Ullio,
  Astrophys.\ J.\  {\bf 566} (2002) 604
  [arXiv:astro-ph/0109463].
  
\bibitem{lines}
 P.~Ullio, L.~Bergstrom, J.~Edsjo and C.~G.~Lacey,
  Phys.\ Rev.\ D {\bf 66} (2002) 123502
  [arXiv:astro-ph/0207125];
  L.~Bergstrom, J.~Edsjo and P.~Ullio,
  Phys.\ Rev.\ Lett.\  {\bf 87} (2001) 251301
  [arXiv:astro-ph/0105048].
\bibitem{BUB} L.~Bergstr\"om,
 P.~Ullio and J.H.~Buckley, Astropart.\ Phys.\ {\bf 9} (1998) 137.

  
\bibitem{glast}
GLAST Proposal to NASA A0-99-055-03 (1999).  
  
\end{thebibliography}
\end{document}